\documentstyle[sprocl]{article}
\input epsf  

\def\be{\begin{equation}}
\def\ee{\end{equation}}
\def\bea{\begin{eqnarray}}
\def\eea{\end{eqnarray}}

\begin{document}

\title{Quark Masses, Chiral Symmetry, and the $U(1)$ Anomaly}
\author{Michael Creutz}
\address{Physics Department,
Brookhaven National Laboratory,
Upton, NY 11973; email:
creutz@bnl.gov}


\maketitle\abstracts{ I discuss the mass parameters appearing in the
gauge theory of the strong interactions, concentrating on the two
flavor case.  I show how the effect of the CP violating parameter
$\theta$ is simply interpreted in terms of the state of the {\ae}ther
via an effective potential for meson fields.  For degenerate flavors I
show that a first order phase transition is expected at $\theta=\pi.$
I speculate on the implications of this structure for Wilson's lattice
fermions.  }
  
\section*{ I. Introduction}

This talk concerns the mass term $m\overline\psi\psi$ in the standard
theory of quarks and gluons.  It is an abridged version
of my recent Phys. Rev. article \cite{chiralmasses}.  
One of my goals is to provide an intuitive picture for the physical
meaning of the CP-violating parameter of the strong interactions.
This term, often called the $\theta$ term, is usually discussed in
terms of topological excitations of the gauge fields.  Here, however,
I treat it entirely in terms of the chiral symmetries expected in the
massless limit of the theory.

I conclude that a first-order transition is expected at $\theta=\pi$
when the flavors have a small degenerate mass.  This transition can be
removed if flavor-breaking is large enough.  At the transition, CP is
spontaneously broken. I will also remark on the implications for the
structure of Wilson's lattice fermions.

This is a subject with a long history, and most of what I say is
buried in numerous previous studies.  The implications of $\theta$ to
the fermion mass matrix are well known to low-energy chiral Lagrangian
discussions \cite{1.}$^{\!-\,}$\cite{8.}.  The possiblity of
a first-order phase transition at large $\theta$ has been discussed in
\cite{2.}.  The possibility of a spontaneous breaking of CP was
pointed out even before the significance of the parameter $\theta$ was
appreciated \cite{3.}.  The relation of $\theta$ to lattice Wilson
fermions was elucidated some time ago by Seiler and Stamatescu
\cite{4.} and was the subject of some recent work of my own
\cite{10.}.

The sign of the fermion mass is sometimes regarded as a convention.
This is indeed the case for ordinary quantum electrodynamics in four
space-time dimensions, where by Furry's theorem \cite{11.} there are
no triangle diagrams and corresponding anomalies.  However, it is
explicitly false for the massive Schwinger model of electrodynamics in
two space-time dimensions \cite{12.,10.}.  Furthermore, as the
remaining discussion will argue, hadronic physics would change if the
sign of one of the quark masses were flipped.

To start the discussion, consider a change of variables
$$
\psi \longrightarrow e^{i\gamma_5\theta/2} \psi.             \eqno (1)
$$
Since $1=(\gamma_5)^2$, this modifies the fermion mass term to
$$
m\overline\psi\psi \longrightarrow m_1\overline\psi\psi
+im_2\overline\psi\gamma_5\psi \eqno (2)
$$
where
$$
\matrix
{m_1 
= m \cos(\theta)\cr
 m_2 
= m \sin(\theta).\cr 
}
\eqno (3)
$$
The kinetic and gauge terms of the quark-gluon action are formally
invariant under this transformation.  Thus, were one to start with the
more general mass term of Eq.~(2), one might expect a physical
situation independent of $\theta$.  However, because of the chiral
anomaly, this is not true.  The angle $\theta$ represents a
non-trivial parameter of the strong interactions.
Its non-vanishing would give rise to CP violating processes.
As such are not observed in hadronic physics, the numerical value of
$\theta$ must be very small \cite{5.}.

\begin{figure}

\epsfxsize .6\hsize
\centerline {\epsfbox{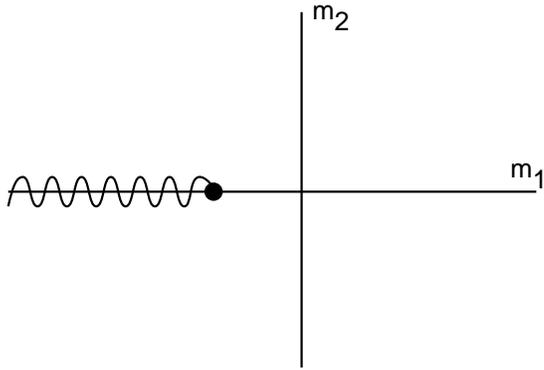}}\bigskip

\caption{The phase diagram for the one flavor case.
The wavy line represents a first-order phase transition, along
which $i\overline\psi\gamma_5\psi$ acquires an expectation value.  The
end point of this transition line is renormalized away from the origin
towards negative $m_1$.}
\end{figure}

If Eq.~(1) just represents a change of variables, how can this affect
physics?  The reason is entwined with the divergences of quantum field
theory and the necessity of regularization.  Fujikawa \cite{13.} has
shown how to incorporate the anomaly into the path integral
formulation via the the fermionic measure, which becomes non-invariant
under the above chiral rotation.  Under a Pauli-Villars \cite{14.}
approach $\theta$ represents a relative $\gamma_5$ rotation between
the mass term for the fundamental particle and the mass term for a
heavy regulator field.  On the lattice with Wilson's fermion
prescription \cite{15.}, the doublers play this role of defining the
relative chiral phase \cite{9.,10.}.

The phase diagram in the $(m_1, m_2)$ plane is strongly dependent on
the number of fermion flavors.  With a single species, a first-order
phase transition line runs down the negative $m_1$ axis, starting at a
non zero value for $m_1$.  This is sketched in Fig.~(1).  For two
flavors I argue for two first-order phase transition lines, starting
near the origin and running up and down the $m_2$ axis.  For
degenerate quarks these transitions meet at the chiral limit of
vanishing fermion mass, while a small flavor breaking can separate the
endpoints of these first-order lines.  This is sketched in Fig.~(2).
With $N_f>2$ flavors, the $(m_1,m_2)$ plane has $N_f$ first order phase
transition lines all pointing at the origin.  The conventionally
normalized parameter $\theta$ is $N_f$ times the angle to a point in
this plane, and these transition lines are each equivalent to $\theta$
going through $\pi$.

\begin{figure}

\epsfxsize .55\hsize
\centerline {\epsfbox{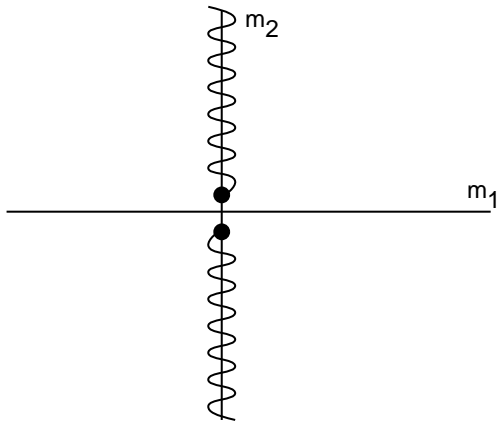}}\bigskip

\caption {The two flavor phase diagram. First-order lines run up and
down the $m_2$ axis.  The second order endpoints of these lines are
separated by a flavor breaking mass difference.  The chiral limit
is pinched between these endpoints.}
\end{figure}

Whenever the number of flavors is odd, there is a first-order
transition running down the negative $m_1$ axis.  Along this line
there is a spontaneous breaking of CP, with a natural order parameter
being $\langle i \overline\psi\gamma_5\psi\rangle$.  This possibility
of a spontaneous breakdown was noted some time ago by Dashen
\cite{3.} and has reappeared at various times in the lattice context
\cite{16.,17.}.

I concentrate my discussion on the two flavor case.  Here several
simplifications make the physics particularly transparent.  I then
discuss how the one flavor result arises when the other flavor is
taken to a large mass.  Finally I conjecture on an analogy with heavy
doublers and Wilson lattice fermions.

\section*{ II. The effective potential}

I begin by defining eight fields around which the discussion
revolves
$$
\matrix
{
\sigma=&c\overline\psi\psi\cr
\vec\pi=&ic\overline\psi\gamma_5\vec\tau\psi\cr
\eta=&ic\overline\psi\gamma_5\psi\cr
\vec\delta=&c\overline\psi\vec\tau\psi.\cr
}\eqno (4)
$$
The fermion $\psi$ has two isospin components, for which $\vec\tau$
represents the standard Pauli matrices.  The factor $c$ is inserted to
give the fields their usual dimensions.  Its value is not particularly
relevant to the qualitative discussion that follows, but one
convention is take $c=F/\vert\langle\overline\psi\psi\rangle\vert$
where $F$ is the pion decay constant and the condensate is in the
standard {\ae}ther.

Corresponding to each of these quantities is a physical spectrum.  In
some cases this is dominated by a known particle.  There is the
familiar triplet of pions around 140 MeV and the eta at 547 MeV.  The
others are not quite so clean, with a candidate for the isoscalar
$\sigma$ being the $f_0(980)$ and for the isovector $\delta$ being the
$a_0(980)$.
I will use that
the lightest particle in the $\delta$ channel appears to be heavier
than the $\eta$.
  
Now consider an effective potential
$V(\sigma,\vec\pi,\eta,\vec\delta)$ constructed for these fields.  I
first consider the theory with vanishing quark masses.  In the
continuum limit, the strong coupling constant is absorbed via the
phenomenon of dimensional transmutation \cite {18.}, and all
dimensionless quantities are in principle determined.  In the full
theory with the quark masses turned back on, the only parameters are
those masses and $\theta$.

For the massless theory many of the chiral symmetries become exact.
Because of the anomaly, the transformation of Eq.~(1), which mixes the
$\sigma$ and $\eta$ fields, is not a good symmetry.  However flavored
axial rotations should be valid.  For example, the rotation
$$
\psi \longrightarrow e^{i\gamma_5\tau_3\phi/2} \psi.             \eqno (5)
$$
mixes $\sigma$ with $\pi_3$ 
$$
\matrix
{
\sigma 
\longrightarrow +\cos(\phi) \sigma + \sin(\phi)\pi_3\cr 
\pi_3  
\longrightarrow -\sin(\phi) \sigma + \cos(\phi)\pi_3\cr 
}
\eqno (6)
$$
This transformation also mixes $\eta$ with $\delta_3$
$$
\matrix
{
\eta     
\longrightarrow +\cos(\phi) \eta + \sin(\phi)\delta_3\cr 
\delta_3 
\longrightarrow -\sin(\phi) \eta + \cos(\phi)\delta_3\cr 
}
\eqno (7)
$$

For the massless theory, the effective potential is invariant under
such rotations.  In this two flavor case, the consequences can be
compactly expressed by going to a vector notation.  I define the four
component objects $\Sigma=(\sigma,\vec\pi)$ and
$\Delta=(\eta,\vec\delta)$.  The effective potential is a function
only of invariants constructed from these vectors.  A complete set of
invariants is $\{\Sigma^2,\ \Delta^2,\ \Sigma\cdot\Delta\}$.  This
separation into two sets of fields is special to the two flavor case,
but makes the behavior of the theory particularly transparent.

\begin{figure}

\epsfxsize .6\hsize
\centerline {\epsfbox{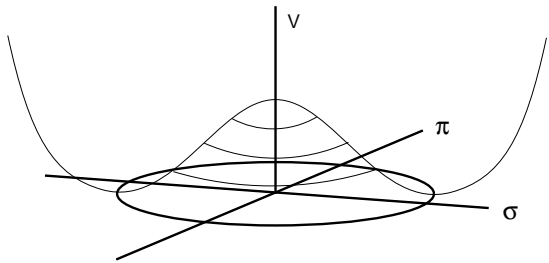}}\bigskip

\caption {The ``sombrero'' potential representing the chiral limit
of massless quarks.}

\end{figure}

I now use the experimental fact that chiral symmetry appears to be
spontaneously broken.  The minimum of the effective potential should
not occur for all fields having vanishing expectation.  We also know
that parity and flavor appear to be good symmetries of the strong
interactions, and thus the expectation value of the fields can be
chosen in the $\sigma$ direction.  Temporarily ignoring the fields
$\Delta$, the potential should have the canonical ``sombrero''
shape, as stereotyped with the form
$$
V=\lambda(\Sigma^2-v^2)^2=\lambda(\sigma^2+\vec\pi^2-v^2)^2 \eqno (8)
$$
Here $v$ is the magnitude of the {\ae}ther expectation value for
$\sigma$, and $\lambda$ is a coupling strength related to the $\sigma$
mass.  The normalization convention mentioned below Eq.~(4) would have
$v=F/2$.  I sketch the generic structure of the potential in
Fig.~(3). This gives the standard picture of pions as Goldstone
bosons associated with fields oscillating along degenerate minima.

Now consider the influence of the fields $\Delta$ on this potential.
Considering small fields, I expand the potential
about vanishing $\Delta$
$$
V=\lambda(\Sigma^2-v^2)^2+\alpha \Delta^2 - \beta
(\Sigma\cdot\Delta)^2+\ldots
\eqno (9)
$$
Being odd under parity, $\Sigma\cdot\Delta$ appears
quadratically.

The terms proportional to $\alpha$ and $\beta$ generate masses for the
$\eta$ and $\delta$ particles.  Since $\Delta^2=\eta^2+\vec\delta^2$,
the $\alpha$ term contributes equally to each.  Substituting
$\Sigma\sim(v,\vec 0)$ gives $(\Sigma\cdot\Delta)^2\sim v^2\eta^2$;
thus, the $\beta$ term breaks the $\eta$--$\vec\delta$ degeneracy.
Here is where the observation that the $\eta$ is lighter than the
$\delta$ comes into play; I have written a minus sign in Eq.~(9), thus
making the expected sign of $\beta$ positive.

Now I turn on the fermion masses.  I consider small masses, and assume
they appear as a general linear perturbation of the effective
potential
$$
V\longrightarrow V-(M_1\cdot\Sigma+M_2\cdot\Delta)/c.
\eqno (10)
$$ 
Here the four-component objects $M_{1,2}$ represent the possible mass
terms.  The normalization constant $c$ appears in Eq.~(4).  The zeroth
component of $M_1$ gives a conventional mass term proportional to
$\overline\psi\psi$, contributing equally to both flavors.  The mass
splitting of the up and down quarks appears naturally in the third
component of $M_2$, multiplying $\overline\psi\tau_3\psi$.  The term
$m_2$ of Eq.~(2) lies in the zeroth component of $M_2$.

The chiral symmetries of the problem now tell us that physics can only
depend on invariants.  For these I can take $M_1^2$, $M_2^2$, and
$M_1\cdot M_2$.  That there are three parameters is reassuring; there
are the quark masses $(m_u, m_d)$ and the CP violating parameter
$\theta$.  The mapping between these parameterizations is non-linear,
the conventional definitions giving
$$
\matrix
{
M_1^2 
&
=& (m_u^2+m_d^2)/4+m_u m_d\cos(\theta)/2\cr
M_2^2 
&
=& (m_u^2+m_d^2)/4-m_u m_d\cos(\theta)/2\cr
M_1\cdot M_2 
&
=& m_u m_d\sin(\theta)/2\cr
}
\eqno (11)
$$
If one of the quark masses, say $m_u$, vanishes, then the $\theta$
dependence drops out.  While this may be a possible way to remove any
unwanted CP violation from the strong interactions, having a single
quark mass vanish represents a fine tuning which is not obviously more
compelling than simply tuning $\theta$ to zero.  Also, having $m_u=0$
appears to be phenomenologically untenable \cite{7.,8.}.

\begin{figure}

\epsfxsize .8\hsize
\centerline {\epsfbox{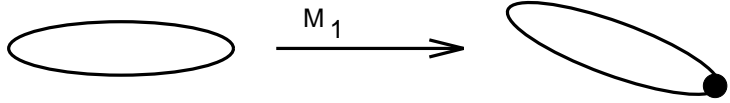}}\bigskip

\caption {The effect of $M_1$ on the effective potential. The ellipse in
this and the following figures represents the minima of the effective
potential from Fig.~(3).  The dot represents where the {\ae}ther settles.}

\end{figure}

I now turn to a physical picture of what the two mass terms $M_1$ and
$M_2$ do to the ``Mexican hat'' structure of the massless potential.
For $M_1$ this is easy; its simply tilts the sombrero.  This is
sketched in Fig.~(4).  The symmetry breaking is no longer spontaneous,
with the tilt selecting the direction for $\Sigma$ field to acquire
its expectation value.  This picture is well known, giving rise to
standard relations such as the square of the pion mass being linearly
proportional to the quark mass \cite{19.}.

The effect of $M_2$ is more subtle.  This quantity has no direct
coupling to the $\Sigma$ field; so, I must look to higher order.  The
$M_2$ term represents a force pulling on the $\Delta$ field, and
should give an expectation value proportional to the strength,
$\langle\Delta\rangle \propto M_2$.  Once $\Delta$ gains an
expectation value, it then effects $\Sigma$ through the $\alpha$ and
$\beta$ terms of the potential in Eq.~(9).  The $\alpha$ term is a
function only of $\Sigma^2$, and, at least for small $M_2$, should not
qualitatively change the structure of the symmetry breaking.  On the
other hand, the $\beta$ term will warp the shape of our sombrero.  As
this term is quadratic in $\Sigma\cdot\Delta$, this warping is
quadratic.  With $\beta$ positive, as suggested above, this favors an
expectation value of $\Sigma$ lying along the vector $M_2$, but the
sign of this expectation is undetermined.  This effect is sketched in
Fig.~(5).

\begin{figure}

\epsfxsize .8\hsize
\centerline {\epsfbox{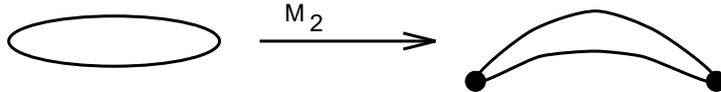}}\bigskip

\caption {The effect of $M_2$ on the effective potential. The dots represent
two places where the {\ae}ther can settle.}

\end{figure}

To summarize, the effect of $M_1$ is to tilt our Mexican hat, while
the effect of $M_2$ is to install a quadratic warping.  The three
parameters of the theory are the amount of tilt, the amount of
warping, and, finally, the relative angle between these effects.  To
better understand the interplay of these various phenomena, I now
consider two specific situations in more detail.

\section*{ III. Case A: $M_1||M_2$}

First consider $M_1$ and $M_2$ parallel in the four vector sense.
This is the situation when we have the two mass terms of Eq.~(2) and
no explicit breaking of flavor symmetry.  Specifically, I take
$M_1=(m_1,\vec 0)$ and $M_2=(m_2,\vec 0)$.  In this case the warping
and the tilting
are along the same axis.

Suppose I consider $m_2$ at some non-vanishing fixed value, and study
the state of the {\ae}ther as $m_1$ is varied.  The $m_2$ term has
warped the sombrero, but if $m_1$ is large enough, the potential will
have a unique minimum in the direction of this pull.  As $m_1$ is
reduced in magnitude, the tilt decreases, and eventually the warping
generates a second local minimum in the opposite $\sigma$ direction.
As $m_1$ passes through zero, this second minimum becomes the lower of
the two, and a phase transition occurs exactly at $m_1=0$.  The
transition is first order since the expectation of $\sigma$ jumps
discontinuously.  This situation is sketched in Fig.~(6).  From
Eq.~(11), the transition occurs at $m_u=m_d$ and $\theta=\pi$.

\begin{figure}

\epsfxsize .8\hsize
\centerline {\epsfbox{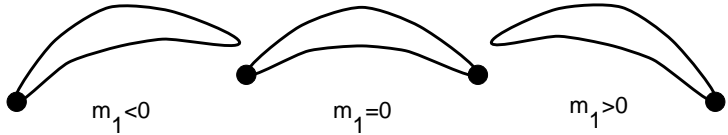}}\bigskip

\caption {Varying $m_1$ at fixed $m_2$.  A first-order phase
transition is expected at $m_1=0$.  This corresponds to $\theta=\pi$.
The dots represent places where the {\ae}ther can settle.}

\end{figure}

As $m_2$ decreases, the warping decreases, reducing the barrier
between the two minima.  This makes the transition softer.  A small
further perturbation in, say, the $\pi_3$ direction, will tilt the
sombrero a bit to the side.  If the warping is small enough, the field
can then roll around the preferred side of the hat, thus opening a gap
separating the positive $m_2$ phase transition line from that at
negative $m_2$.  In this way sufficient flavor breaking can remove the
first-order phase transition at $\theta=\pi$.  If I start at
$\theta=0$ with a mass splitting between the up and down quarks, an
isoscalar chiral rotation to non-zero $\theta$ will generate
just such a term.

\section*{ IV. Case B: $M_1\perp M_2$}

I now turn to a situation where $M_1$ and $M_2$ are orthogonal.  To be
specific, take $M_1=(m_1,\vec 0)$ and $M_2=(0,0,0,\delta m)$, which
physically represents a flavor symmetric mass term $m_1=(m_u+m_d)/2$
combined with a flavor breaking $\delta m=(m_u-m_d)/2$.  Now $M_2$
warps the sombrero downwards in the $\pm \pi_3$ direction.  A large
$m_1$ would overcome this warping, still giving an {\ae}ther with only
$\sigma$ having an expectation value.  However, as $m_1$ decreases in
magnitude with a fixed $\delta m$, there eventually comes a point
where the warping dominates the tilting.  At this point we expect a
new symmetry breaking to occur, with $\pi_3$ acquiring an expectation
value.  This is sketched in Fig.~(7).  As $\pi_3$ is a CP odd
operator, this is a spontaneous breaking of CP.

To make this into a proper two dimensional phase diagram, I add an
$m_3\pi_3$ piece to the potential.  This effectively twists $M_1$ away
from being exactly perpendicular to $M_2$.  
This term
explicitly breaks CP and can be expected to
remove the transition, just as an applied field removes the phase
transition in the Ising model.  We thus have a phase diagram in the
$(m_1,m_3)$ plane with a first-order transition connecting two
symmetrically separated points on the $m_1$ axis.  This is sketched in
Fig.~(8).

\begin{figure}

\epsfxsize .8\hsize
\centerline {\epsfbox{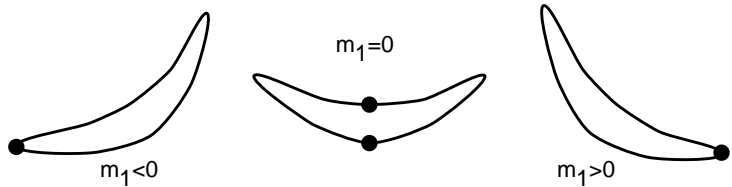}}\bigskip

\caption {Varying $m_1$ at fixed quark mass splitting.  A second order
phase transition occurs when the tilting is reduced sufficiently for a
spontaneous expectation of $\pi_3$ to develop.  The dots represent
places where the {\ae}ther can settle.}

\end{figure}

Physically, the endpoints of this transition line are associated with
the points where the respective quark masses vanish.  The phase
transition occurs when the two flavors have masses of opposite sign.
Simultaneously flipping the signs of both quark masses can always be
done by a flavored chiral rotation, say about the $\pi_3$ axis, and
thus is a good symmetry of the theory.

\begin{figure}

\epsfxsize .7\hsize
\centerline {\epsfbox{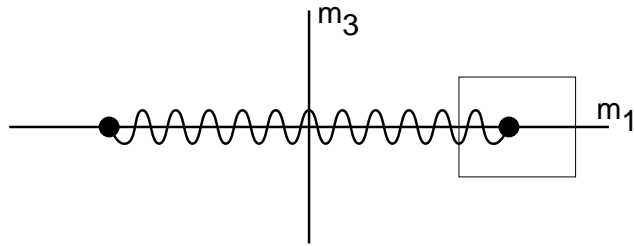}}\bigskip

\caption {The $(m_1,m_3)$ phase diagram for unequal mass quarks.  The
wavy line represents a first-order phase transition ending at the
second order dots.  The light box on the right shows how the one
flavor diagram of Fig.~(1) is extracted.}

\end{figure}

Taking one of the flavors to infinite mass provides a convenient way
to understand the one flavor situation.  As sketched in Fig.~(8), this
represents looking only at the vicinity of one endpoint of the
transition line.  In terms of the light species, this transition
represents a spontaneous breaking of CP with a non-vanishing
expectation for $i\overline\psi\gamma_5\psi$.  In the lattice context
the possibility of such a phase was mentioned briefly by Smit \cite{16.},
and extensively discussed by Aoki and Gocksch \cite{17.}.

\section*{ VI. Implications for Wilson's lattice fermions}

The Lagrangian for free Wilson lattice fermions is \cite {15.}
$$
\matrix
{
L(K,r,M)=\cr
\sum_{j,\mu}
     K\left(\overline\psi_j         ( i\gamma_\mu+r) \psi_{j+e_\mu}
           +\overline\psi_{j+e_\mu} (-i\gamma_\mu+r) \psi_{j}\right)\cr
+\sum_{j} (m_1 \overline\psi_j\psi_j+i m_2 \overline\psi_j\gamma_5\psi_j) \cr
}
\eqno (12)
$$
Here $j$ labels the sites of a four dimensional hyper-cubic lattice,
$\mu$ runs over the space time directions, and $e_\mu$ is the unit
vector in the $\mu$'th direction.  I have scaled out all factors of
the lattice spacing.  The parameter $K$ is called the hopping
parameter, and $r$ controls the strength of the so called ``Wilson
term,'' which separates off the famous doublers.  I have also added
an unconventional $m_2$ type mass term to
connect with my earlier discussion.

Being quadratic with only nearest neighbor couplings, the spectrum is
easily found by Fourier transformation.  Conventionally, a massless
fermion is obtained by taking $m_1=8Kr$, but there are other places
where this original particle is massive while other doublers from the
naive theory become massless.  At $m_1=-8Kr$ one such species does so,
at each of $m_1=\pm 4Kr$ there are four massless doublers, and at
$m_1=0$ I find the remaining $6$ of the total 16 species present in
the naive theory.

I conjecture that these various species should be thought of as
flavors.  When the gauge fields are turned on, then the full chiral
structure should be a natural generalization of the earlier
discussion.  Thus near $m_1=8Kr$ I expect a first-order transition to
end, much as is indicated in Fig.~(1).  This may join with numerous
other transitions at the intermediate values of $m_1$, all of which
then finally merge to give a single first-order transition line ending
near $m_1=-8Kr$.  The situation near $0$ and $\pm 4Kr$ involves larger
numbers of flavors, and properly requires 
a more general analysis.
One possible way the lines could join up is shown
in Fig.~(9a).

\begin{figure}

\epsfxsize .75\hsize
\centerline {\epsfbox{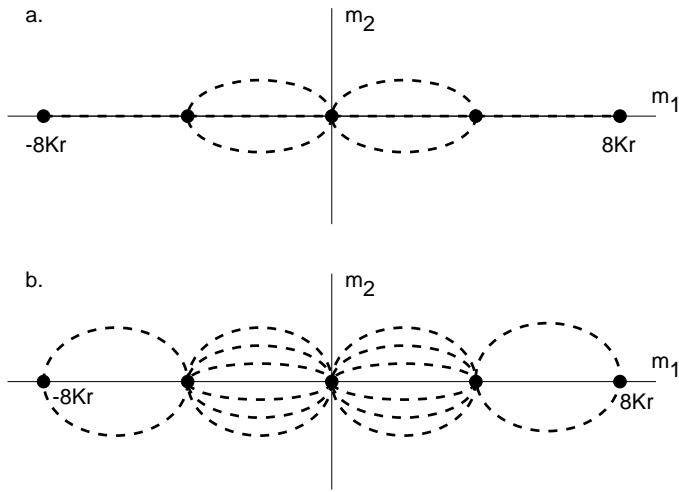}}\bigskip

\caption {Possible phase diagrams for lattice gauge theory with Wilson
fermions.  The dashed lines represent first-order phase transitions
and the dots represent points where massless excitations should exist.
Parts (a) and (b) are for the one and two flavor cases, respectively.}

\end{figure}

For two flavors of Wilson fermions, if we look near to the singularity
at $8Kr$ we should obtain a picture similar to Fig.~(2).  However,
further away these lines can curve and eventually end in the
structure at the other doubling points.  One possible picture is
sketched in Fig.~(9b).

\section*{ VIII. Summary and conclusions}

I have presented a physical picture of the parameter $\theta$ in the
context of an effective potential for spin-zero bilinears of quark
fields.  I have argued for a first-order transition at $\theta=\pi$
when all flavors are degenerate, and shown how flavor breaking can
remove this transition.

A number of years ago Tudron and I \cite{22.} conjectured on the
interplay of the confinement mechanism with $\theta$, and speculated
that confinement might make $\theta$ unobservable.  Recently
Schierholz \cite{23.}  argued that keeping confinement in the
continuum limit may drive the theory to $\theta=0$.  The connection
with present discussion is unclear, but the symmetries seem to
indicate no obvious problem with $\theta$ being observable.
Furthermore, the fact that the $\eta$ is lighter than particle
candidates in the $\delta$ channel suggests that there indeed must be
the $\beta$ term of Eq.~(9), and it is this term which is directly
responsible for the physical dependence on $\theta$.

\eject


\section*{References}
\begin{thebibliography}{99}

\bibitem{chiralmasses} M.~Creutz, Phys.~Rev.~D52, 2951 (1995).

\bibitem{1.} G.~'t Hooft, Phys.~Rev.~Lett. 37, 8 (1976); 
  Phys.~Rev. D14 (1976) 3432.

\bibitem{2.} E. Witten, Annals of Phys. 128, 363 (1980).


\bibitem{3.} R. Dashen, Phys. Rev. D3, 1879 (1971).  

\bibitem{4.}  J. Gasser and H. Leutwyler, Ann. Phys. 158, 142 (1984);
H. Leutwyler, Ann. Phys. 235, 165 (1994); lectures at workshop
``Hadrons 1994,'' Gramado, RS, Brazil (1994) (HEP-PH-9406283).

\bibitem{5.} V. Baluni, Phys. Rev. D19, 2227 (1979);
R. Crewther, P. Di Vecchia, G. Veneziano, and I. Witten,
Phys. Lett. 88B, 123 (1979).

\bibitem{6.} E. Witten, Nucl. Phys. B223, 422 (1983).

\bibitem{7.} J. Donoghue, B. Holstein, and D. Wyler,
Phys. Rev. Lett. 69, 3444 (1992); H. Leutwyler, Nucl. Phys. B337 108
(1990).
 
\bibitem{8.} J. Bijnens, J. Prades, and E. de Rafael,
Phys.Lett. B348 226-238 (1995). 

\bibitem{9.} E.~Seiler and I.~Stamatescu, Phys.~Rev. D25 (1982) 2177.

\bibitem{10.} M. Creutz, Nucl. Phys. B (Proc. Suppl.) 42, 56 (1995).

\bibitem{11.} W. Furry, Phys. Rev. 81, 115 (1937).

\bibitem{12.} S. Coleman, Annals of Phys. 101, 239 (1976).

\bibitem{13.} K. Fujikawa, Phys. Rev. Lett. 42, 1195 (1979).

\bibitem{14.} W. Pauli and F. Villars, Rev. Mod. Phys. 21, 433 (1949).

\bibitem{15.} K.~Wilson, in {\sl New Phenomena in Subnuclear Physics}, 
  Edited by A. Zichichi (Plenum Press, NY, 1977), p. 24. 

\bibitem{16.} J.~Smit, Nucl.~Phys. B175 307 (1980).

\bibitem{17.} S. Aoki, Nucl. Phys. B314, 79 (1989); S.~Aoki and A.~Gocksch, 
Phys.~Rev. D45, 3845 (1992).

\bibitem{18.} S. Coleman and E. Weinberg, Phys. Rev. D7 1888 (1973).

\bibitem{19.} M. Gell-Mann, R. Oakes, and B. Renner, Phys. Rev. 175,
2195 (1968).

\bibitem{20.} S.~Aoki, S.~Boetcher, and A.~Gocksch, Phys.~Lett. B331, 157
  (1994); K.~Bitar and P.~Vranas, Phys.~Rev. D50, 3406 (1994); Nucl.~Phys.
  B, Proc.~Suppl. 34, 661 (1994).

\bibitem{21.} M. Creutz, Phys. Rev. Lett. 46, 1441 (1981); M. Creutz
and K.J.M. Moriarty, Phys. Rev. D25, 1724 (1982).

\bibitem{22.} M.~Creutz and T.~Tudron, Phys.~Rev. D16 2978 (1977).

\bibitem{23.} G.Schierholz, Nucl.~Phys.  B, Proc.~Suppl. 42, 270
(1995).

\end{thebibliography}
\end{document}